\documentclass[a4paper]{article}
\usepackage{Odyssey2022}
\usepackage{epsfig,amssymb,amsmath}
\usepackage{hyperref}
\usepackage{booktabs}
\usepackage{array}
\newcolumntype{P}[1]{>{\centering\arraybackslash}p{#1}}
\newcolumntype{M}[1]{>{\centering\arraybackslash}m{#1}}
\newcolumntype{R}[1]{>{\raggedleft\arraybackslash}p{#1}}
\usepackage{bm}
\usepackage{color}
\usepackage{multirow}
\usepackage{tablefootnote}
\usepackage{makecell}

\usepackage{xcolor}

\def\S{{\mathcal{S}}}

\ninept

\title{A Probabilistic Fusion Framework for Spoofing Aware Speaker Verification
}

\name{You Zhang, Ge Zhu, Zhiyao Duan}

\address{Audio Information Research Lab,
University of Rochester, Rochester, NY, USA \\
{\small \tt you.zhang@rochester.edu} }
\begin{document}
\maketitle

\begin{abstract}
The performance of automatic speaker verification (ASV) systems could be degraded by voice spoofing attacks. Most existing works aimed to develop standalone spoofing countermeasure (CM) systems. Relatively little work targeted at developing an integrated spoofing aware speaker verification (SASV) system. In the recent SASV challenge, the organizers encourage the development of such integration by releasing official protocols and baselines. In this paper, we build a probabilistic framework for fusing the ASV and CM subsystem scores. We further propose fusion strategies for direct inference and fine-tuning to predict the SASV score based on the framework. Surprisingly, these strategies significantly improve the SASV equal error rate (EER) from 19.31\% of the baseline to 1.53\% on the official evaluation trials of the SASV challenge. We verify the effectiveness of our proposed components through ablation studies and provide insights with score distribution analysis.
\end{abstract}

\section{Introduction}
Automatic speaker verification (ASV) aims to verify the identity of the target speaker given a test speech utterance. A typical speaker verification process involves two stages: First, a few utterances of the speaker are enrolled, then the identity information extracted from the test utterance is compared with that of the enrolled utterances for verification~\cite{Hannani2009text}. ASV researchers have been developing speaker embedding extraction methods~\cite{tu21_interspeech, zhu21b_interspeech, zhu21c_interspeech} to encode speaker identity information for verification. However, it is likely that the test utterance is not human natural speech but \textit{spoofing attacks} that try to deceive the ASV system. Spoofing attacks usually include impersonation, replay, text-to-speech, voice conversion attacks. Studies have shown that ASV systems are vulnerable to spoofing attacks~\cite{WU2015130}. 

In recent years, researchers have been developing spoofing countermeasure (CM) and audio deepfake detection systems to detect spoofing attacks. 
With the ASVspoof 2019 challenge which provides a large-scale standard dataset and evaluation metrics, the CM systems have been improved in various aspects, especially on the generalization ability~\cite{chen2020generalization, zhang21one, cheng21b_interspeech} and channel robustness~\cite{chen21_asvspoof, kang21b_asvspoof, wang2019cross} for in-the-wild applications. However, 
all of the above works focused on the evaluation of standalone CM systems. Intuitively, an imperfect CM system would accept spoofing attacks but reject bona fide speech from the target person~\cite{Sahidullah+2016}.
After all, the ultimate goal of developing a CM system is to protect the ASV system from falsely accepting spoofing attacks. However, how an improved CM system benefits the ASV system is not clear. Although the minimum t-DCF~\cite{kinnunen2018t} used in the ASVspoof challenge~\cite{nautsch2021asvspoof} evaluates the reliability of CM systems to ASV systems, it is calculated on a fixed ASV system provided by the ASVspoof organizers instead of being adapted to the ASV system at hand. For better protecting the ASV system from being spoofed and maintaining its discrimination ability on speaker identity, the CM component should be jointly optimized with the ASV system. As a result, an integrated ASV and CM system is promising.


Relatively little attention is paid to improving the integration of ASV and CM systems. As reviewed in Section~\ref{sec:liter}, some work has proposed some frameworks to address such problem, but due to the lack of standard metrics and datasets, it is hard to benchmark the state-of-the-art spoofing aware speaker verification (SASV) system. Recently, the SASV challenge~\cite{jung2022sasv} has been held to further encourage the study of integrated systems of ASV and CM. In this challenge, only cases of logical access (LA) spoofing attacks, i.e., TTS and VC attacks, are taken into consideration. The test utterances of the SASV system can be categorized into three classes: \textit{target}---bona fide speech belonging to the target person, \textit{non-target}---bona fide speech but not belonging to the target speaker, and \textit{spoof}---spoofing attacks. 

In this work, we formulate a fusion-based SASV system under the probabilistic  framework on top of the ASV and CM subsystems. We also propose a fine-tuning strategy on the integrated system for further improvement. With the proposed fusion strategies, we outperform the SASV baseline systems by a large margin. Our best performing system achieved 1.53\% SASV-EER on the official evaluation trials. We also provide an ablation study and score distribution analysis for future study.


\begin{figure}[]
  \centering
  \centerline{\includegraphics[width=1\linewidth]{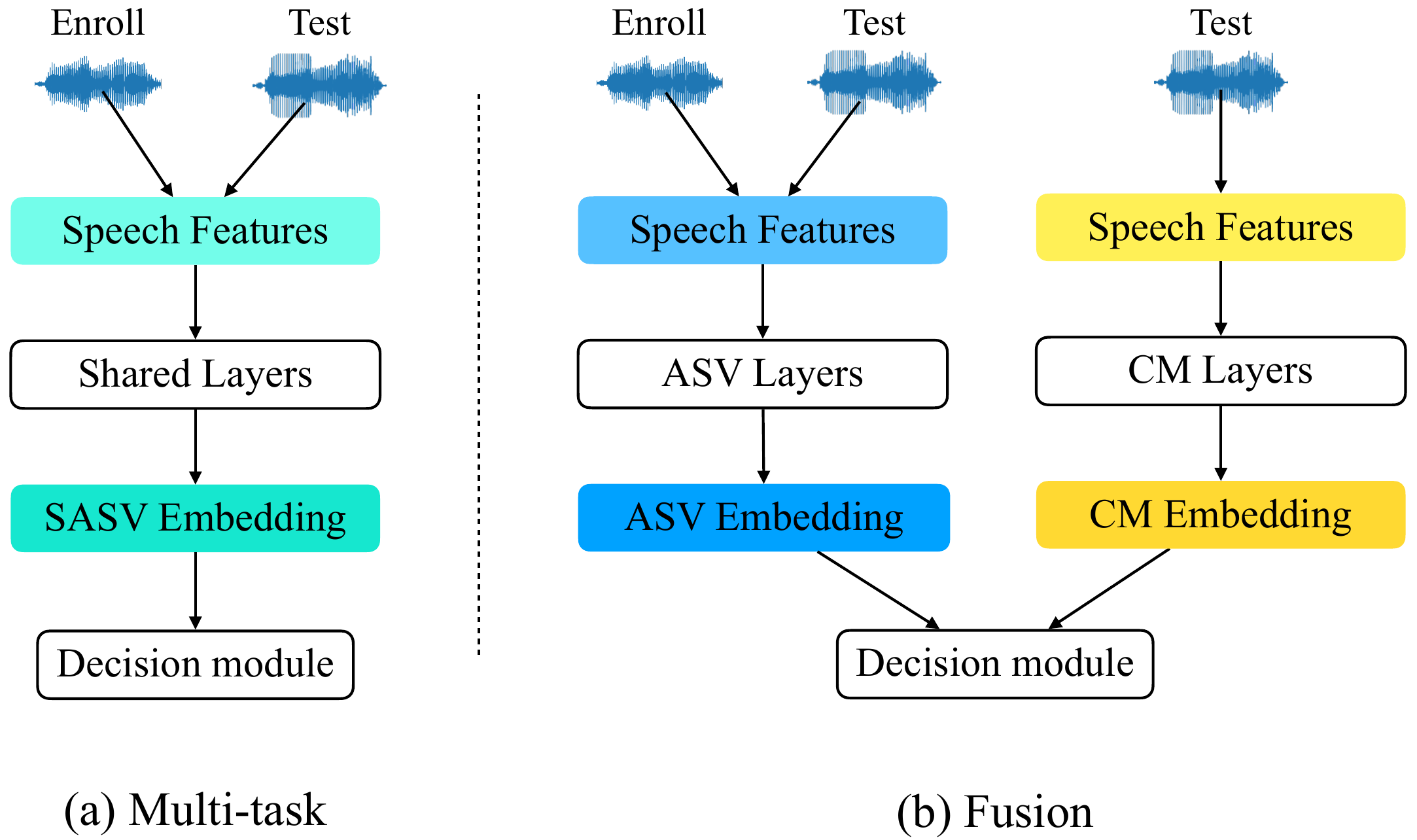}}
\caption{Illustration of two categories of methods in the literature of SASV systems. The ``layers'' represent different machine learning models aiming to extract embeddings such as $i$-vectors. The ``decision module'' could be (i) a layer for computing the final score on the SASV embedding, or (ii) a score fusion module that fuses ASV and CM scores.
}
\label{fig:literature}
\end{figure}

\section{Literature review}
\label{sec:liter}
In the literature, the SASV system is usually referred to as joint ASV and CM systems. There are mainly two categories of methods: (a) multi-task learning-based and (b) fusion-based. The comparison of their general structures is illustrated in Fig.~\ref{fig:literature}. 


\subsection{Multi-task learning-based methods}
Li et al.~\cite{li2019multi} proposed a SASV system to perform a joint decision by multi-task learning. The ASV task and CM task share the same spectrum features and a few network layers. A three-stage training paradigm with pre-training, re-training, and speaker enrollment is proposed to extract a common embedding and perform classification with separate classifiers for the two sub-tasks. 
They further extended their work in~\cite{li2020joint} by training the common embedding with triplet loss and then using probabilistic linear discriminant analysis (PLDA) scoring for inference.
Zhao et al.~\cite{zhao2020multi} adapt the multi-task framework with max-feature map activation and residual convolutional blocks to extract discriminative embeddings.

The training of such multi-task neural networks requires both the speaker label and the spoofing labels, so they are trained on ASVspoof datasets which have a limited number of speakers. This might lead the model to overfit the seen speakers and limit their performance in real-world applications. 

\subsection{Fusion-based methods}
As shown in Fig.~\ref{fig:literature}(b), independent ASV and CM models extract separate embeddings to make a joint decision. The speaker (SPK) embedding aims to encode the identity information. The CM embedding is usually the output from the second last layer in the anti-spoofing network.

Some methods perform fusion in the embedding space. Sizov et al.~\cite{sizov2015joint} proposed a two-stage PLDA method for optimizing the joint system in the i-vector space. First, it trains a simplified PLDA model using only the embeddings of the bona fide speech. Then, it estimates a new mean vector, adds a spoofing channel subspace, and trains it using only the embeddings of the spoofed speech.
Gomez et al.~\cite{gomez2020joint} proposed an integration framework with fully connected (FC) layers following the concatenated speaker and CM embeddings.

Some methods perform fusion in the score level. The ASV score is usually the cosine similarity between the speaker embeddings of the enrollment utterances and test utterances. The CM score is the final output of the anti-spoofing model.
Sahidullah et al.~\cite{Sahidullah+2016} first studied the cascade and parallel integrations of ASV with CM to combine scores.
Todisco et al.~\cite{Todisco2018} proposed a Gaussian back-end fusion method that fuses the scores with log-likelihood ratio according to separately modeled Gaussian mixtures. 
Kanervisto et al.~\cite{kanervisto2021optimizing} proposed a reinforcement learning paradigm to optimize tandem detection cost function (t-DCF) by jointly training a tandem ASV and CM system.
Shim et al.~\cite{shim2020integrated} proposed a fusion-based approach that takes the speaker embedding and CM prediction as input and weighs the ASV score, CM score, and their multiplication to make the final decision. 

\begin{figure}[]
  \centering
  \centerline{\includegraphics[width=1\linewidth]{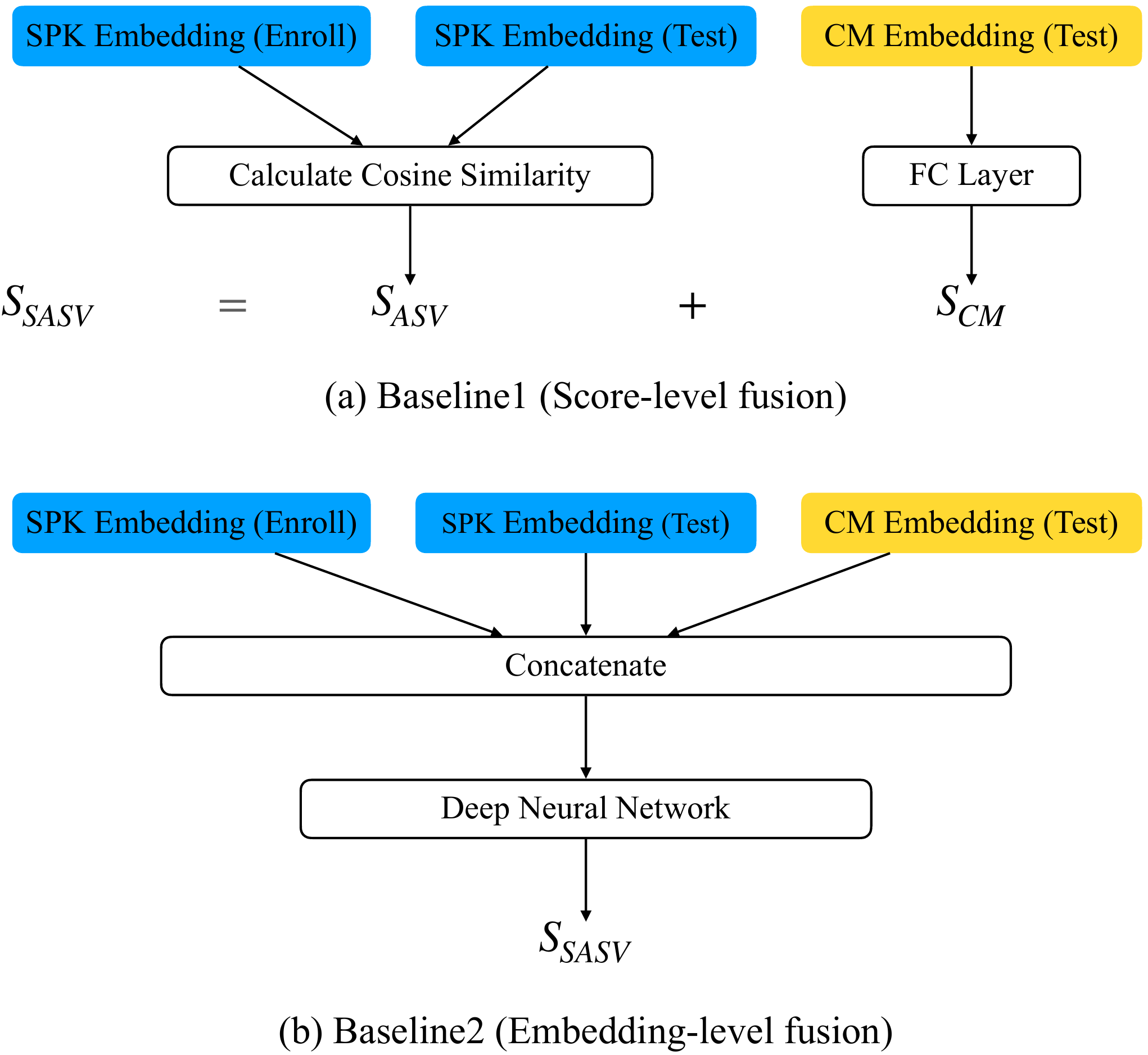}}
\caption{Model structure of the baseline methods from the SASV challenge. Colored boxes denote the embeddings and the bordered boxes represent the operations.
}
\label{fig:baselines}
\end{figure}

\textbf{SASV Baseline methods.}
The SASV challenge~\cite{jung2022sasv} introduces two baselines built upon pre-trained state-of-the-art ASV and CM systems. The structure of the two methods is shown in Fig.~\ref{fig:baselines}.
Baseline1 is a score-level fusion method that sums the scores produced by the separate systems. There is no training involved.
Besides, Baseline2 is an embedding-level fusion method that trains a deep neural network based on concatenated embeddings. The pre-trained speaker and CM embeddings are fixed during training the deep neural network. This is similar to the method proposed in~\cite{gomez2020joint}.

\section{Method}
\subsection{Problem formulation}
Given an enroll utterance $u^{e}$ and a test utterance $u^{t}$, SASV systems need to classify $u^{t}$ into $y^t \in \{0, 1\}$, where 1 represents \textit{target} and 0 includes both \textit{non\textendash target} and \textit{spoof}. In this paper, we focus on a fusion-based SASV system consisting of a pre-trained ASV subsystem and a pre-trained CM subsystem. 
In fusion-based SASV systems, The ASV subsystem computes speaker embeddings $x^{e}_\textit{ASV}$ for the enrollment utterance $u^{e}$ and $x^{t}_\textit{ASV}$ for the test utterance $u^{t}$.
The CM subsystem computes the CM embedding $x^{t}_\textit{CM}$ for $u^{t}$. 
We use pre-trained embedding methods for the ASV subsystem~\cite{Desplanques2020} and the CM subsystem~\cite{Jung2022AASIST}, as they both achieve state-of-the-art discrimination abilities on their respective tasks.

As it is a binary classification problem, 
we use the posterior probability that the test utterance belongs to the positive class (i.e., the \textit{target} class), conditioned on the speaker embeddings, as the final decision score $\S_\textit{SASV}$.
\begin{equation}\label{eq:score}
    \S_\textit{SASV} = P(y^t = 1 |x^{e}_\textit{ASV},x^{t}_\textit{ASV},x^{t}_\textit{CM}).
\end{equation}

For score-level fusion methods, the ASV and CM subsystems each computes a decision score. Similar to Eq.~\eqref{eq:score}, such decision scores can be defined as the posterior probabilities, as $P(y^t_\textit{ASV} = 1 |x^{e}_\textit{ASV},x^{t}_\textit{ASV})$ and $P(y^t_\textit{CM} = 1 |x^{t}_\textit{CM})$, respectively.
Here $y_\textit{ASV}^t$ and $y_\textit{CM}^t \in \{0, 1\}$ are the underlying ground-truth labels along the ASV and CM aspects, respectively. In other words, $y_\textit{ASV}^t = 1$ and $y_\textit{ASV}^t = 0$ indicate that the test utterance is target and non-target, respectively. $y_\textit{CM}^t = 1$ and $y_\textit{CM}^t = 0$ indicate that the test utterance is bona fide and spoof, respectively.

It is noted that these definitions of scores using posterior probabilities are different from those in the baseline methods in Figure~\ref{fig:baselines}. There $\S_\textit{ASV}$ is defined as the cosine similarity between the enrollment embedding and the test embedding, and $\S_\textit{CM}$ is defined as the output of an FC layer. Both value ranges are not between 0 and 1. In the following, we will propose ways to revise the scores in Figure~\ref{fig:baselines}(a) to fit into the proposed probabilistic framework.

\subsection{Probabilistic framework}

We propose a probabilistic framework based on product rule (PR) inspired by~\cite{kittler1998combining}. By definition, $y^t=1$, i.e., the test utterance is \textit{target}, if and only if $y^t_\textit{ASV}=1$ and $y^t_\textit{CM}=1$. Therefore, assuming conditional independence between $y_\textit{ASV}^t$ and $y_\textit{CM}^t$ on the speaker embeddings, we have
\begin{equation}
    \begin{aligned}
    &P(y^t=1|x^{e}_\textit{ASV},x^{t}_\textit{ASV},x^{t}_\textit{CM})\\= &P(y^t_\textit{ASV}=1,y^t_\textit{CM}=1|x^{e}_\textit{ASV},x^{t}_\textit{ASV},x^{t}_\textit{CM})\\
    = &P(y^t_\textit{ASV}=1|x^{e}_\textit{ASV},x^{t}_\textit{ASV},x^{t}_\textit{CM}) P(y^t_\textit{CM}=1|x^{e}_\textit{ASV},x^{t}_\textit{ASV},x^{t}_\textit{CM})\\
    = &P(y^t_\textit{ASV}=1|x^{e}_\textit{ASV},x^{t}_\textit{ASV}) P(y^t_\textit{CM}=1|x^{t}_\textit{CM}).
    \end{aligned}
\label{eq:pr}
\end{equation}
The last equation follows from the fact that $y^t_{ASV}$ is independent from $x^t_{CM}$ and that $y^t_{CM}$ is independent from $x^e_{ASV}$ and $x^t_{ASV}$, as we use pre-trained ASV and CM subsystems. If however, such subsystems are fine tuned during the SASV task, as in Section~\ref{sssec:finetune}, this independence will not be valid anymore.

\subsection{Proposed strategies}




\subsubsection{Direct inference strategy}
\label{sssec:infer}
We adopt the same model structure as the base of the Baseline1 method, shown in Fig.~\ref{fig:baselines} (a). The ASV subsystem outputs the cosine similarity 
between the speaker embedding $x^{e}_\textit{ASV}$ and $x^{t}_\textit{ASV}$. The CM system outputs the CM score $\S_\textit{CM}$ from an FC layer. As both the ASV and CM subsystems are pre-trained and there is no fine tuning in any part of the entire system, this is a direct inference strategy.

As mentioned above, both the ASV score and the CM score do not fit to the proposed probabilistic framework. Therefore, we propose ways to modify their value range to $[0,1]$.
The CM subsystem was pre-trained with a softmax binary classification loss, so the output score $\S_\textit{CM}$ after a sigmoid function $\sigma(x)$ would naturally fit to the range of $[0,1]$, therefore, we define
\begin{equation}
\label{eq:p_cm}
    P(y^t_\textit{CM}=1|x^{t}_\textit{CM}) = \sigma(\mathcal{S}_\textit{CM}).
\end{equation}
For the ASV score, we need some function $f$ to monotonically map the cosine similarity score to a value between 0 and 1:
\begin{equation}
\label{eq:p_asv}
    P(y^t_\textit{ASV}=1|x^{e}_\textit{ASV},x^{t}_\textit{ASV}) = f(\mathcal{S}_\textit{ASV}),
\end{equation}
where $f$ can be a hand-crafted function or some data-driven mapping. Combining Eq.\eqref{eq:score}-\eqref{eq:p_asv}, the final decision score for SASV is represented as:
\begin{equation}
\label{eq:pr_s}
    \S_\textit{SASV} = \sigma(\mathcal{S}_\textit{CM}) \times f(\mathcal{S}_\textit{ASV}).
\end{equation}

By varying the function $f$, we propose three systems using the direct inference strategy. A straightforward method is through a linear mapping $f(s) = (s+1)/2$. We refer to this system as \texttt{PR-L-I}, where \texttt{L} stands for ``linear'' and \texttt{I} is short for ``inference''. For non-linear mapping, we choose the sigmoid function and denote the system as \texttt{PR-S-I}, where \texttt{S} means ``sigmoid''. A potential advantage of a sigmoid function over the linear mapping is that it expands the data range around 0, the more ambiguous region for decisions.
It is noted that neither the linear or sigmoid mapping can result in probabilities that follow the true posterior distribution, therefore, we introduce a third mapping that is trained on the bona fide trials of the development set for $\mathcal{S}_\textit{ASV}$. 
To be specific, we sample target and non-target trials and train a calibration function with logistic regression~\cite{Brummer06focal}, where the target class is map to 1 and the non-target class is mapped to 0. This can be viewed as a data-driven score calibrator. This system using the data-driven calibrated ASV score is represented as \texttt{PR-C-I}. It is expected that when the test utterance is drawn from the same distribution of the trials used to train the calibrator, the ASV subsystem performance would be improved. This hypothesis is verified in our experiments in Table~\ref{tab:main}.



\subsubsection{Fine-tuning strategy}
\label{sssec:finetune}
When the ASV and CM subsystems are fine tuned on the SASV task, then the conditional independence assumption in the last equality of Eq.~\eqref{eq:pr} no longer holds. Instead, we can have an alternative derivation of the posterior probability:
\begin{equation}
    \begin{aligned}
    &P(y^t=1|x^{e}_\textit{ASV},x^{t}_\textit{ASV},x^{t}_\textit{CM})\\=& P(y^t_\textit{ASV}=1, y^t_\textit{CM}=1|x^{e}_\textit{ASV},x^{t}_\textit{ASV},x^{t}_\textit{CM})\\
    =& P(y^t_\textit{ASV}=1|x^{e}_\textit{ASV},x^{t}_\textit{ASV},x^{t}_\textit{CM}) P(y^t_\textit{CM}=1|y^t_\textit{ASV}, x^{e}_\textit{ASV},x^{t}_\textit{ASV},x^{t}_\textit{CM})\\
    =& P(y^t_\textit{ASV}=1|x^{e}_\textit{ASV},x^{t}_\textit{ASV}) P(y^t_\textit{CM}=1|y^t_\textit{ASV}, x^{t}_\textit{CM}).
    \end{aligned}
\label{eq:pr_t}
\end{equation}
The second equality is based on the chain rule and it treats $y^t_\textit{ASV}$ as a condition. It can be interpreted as that the prediction of the CM subsystem depends on that of the ASV subsystem. This dependency can be realized through fine-tuning the CM subsystem conditioned on the ASV system's output score. To do so, 
we fine-tune the FC layer of the CM subsystem while keeping the ASV score fixed in Figure~\ref{fig:baselines}(a).
Instead of fitting $\S_\textit{CM}$ with CM labels, our model directly optimizes the joint score. The training is based on the ground-truth label of whether the test utterance belongs to the \textit{target} class. In other words, the \textit{spoof} and \textit{non-target} utterances share the same negative labels. The final decision score $\mathcal{S}_\textit{SASV}$ is calculated with Eq.~\eqref{eq:pr_s}. 

We fine-tune the system with a prior-weighted binary cross-entropy loss for $\mathcal{S}_\textit{SASV}$. 
The ASV embedding network is pre-trained and fixed, hence the ASV score $\mathcal{S}_\textit{ASV}$ is fixed. Only the FC Layer on top of the CM embedding network is trained and the CM score $\mathcal{S}_\textit{CM}$ is adjusted. During back-propagation, thanks to the multiplication, the gradient of the CM score with respect to the parameters in the FC layer is weighted based on the scaled ASV scores. The gradient receives a larger weight for larger $\mathcal{S}_\textit{ASV}$, which corresponds to utterances that are more similar to the target speaker. This helps the model to pay more attention to such more difficult samples, manifesting an idea of \textit{speaker-aware anti-spoofing}. 

In fine tuning strategy, we choose $f$ as the linear or the sigmoid function, denoted as \texttt{PR-L-F} and \texttt{PR-S-F} respectively. \texttt{L} and \texttt{S} represent the two mapping functions as in Section~\ref{sssec:infer}, while \texttt{F} is short for ``fine-tuning''. We discard the calibration method to prevent over-fitting on the trials dataset.

\section{Experimental setup}

\subsection{Dataset}
\label{ssec:dataset}
ASVspoof 2019 LA~\cite{WANG2020101114} is a standard dataset designed for the LA sub-challenge of ASVspoof 2019. It consists of bona fide speech and a variety of TTS and VC spoofing attacks. The bona fide speech is collected from the VCTK corpus~\cite{yamagishi2019cstr}, while the speakers are separated into three subsets: training (Train), development (Dev), and evaluation (Eval). The spoofed speech in each subset is targeted to spoof the corresponding speakers. The algorithms for spoofing attacks in the evaluation set are totally different from those in the Train and Dev sets. The non-overlap is designed to encourage the generalization ability to unseen attacks for CM systems. Details are shown in Table~\ref{tab:dataset}.

\begin{table}[]
\renewcommand{\arraystretch}{1.2}
\centering
\footnotesize
\caption{Summary of the ASVspoof 2019 LA dataset.}
\begin{tabular}{c|c|c|cc}
\hline\hline
     \multirow{2}{*}{Partition} & \multirow{2}{*}{\#speakers} & Bona fide & \multicolumn{2}{c}{Spoofing attacks} \\ \cline{3-5} 
      &    & \#utterances & \multicolumn{1}{c|}{\#utterances} & Attacks type \\ \hline
Train & 20 & 2,580        & \multicolumn{1}{c|}{22,800}       & A01 - A06    \\
Dev   & 20 & 2,548        & \multicolumn{1}{c|}{22,296}       & A01 - A06    \\
Eval  & 67 & 7,355        & \multicolumn{1}{c|}{63,882}       & A07 - A19    \\
\hline\hline
\end{tabular}
\label{tab:dataset}
\end{table}

For the SASV challenge, the organizers provided official development and evaluation protocols listing the \textit{target}, \textit{non-target}, and \textit{spoof} trials based on the ASVspoof 2019 LA dataset. For each test trial, there are multiple corresponding enrollment utterances to register the target speaker.

\subsection{Evaluation metrics}
Equal error rate (EER) is widely used for binary classification problems, especially in speaker verification and anti-spoofing. It is calculated by setting a threshold such that the miss rate is equal to the false alarm rate. The lower the EER is, the better the discriminative ability has the binary classification system.

SASV-EER is used as the primary metric to evaluate the SASV performance. The SV-EER and SPF-EER are auxiliary metrics to assess the performance of ASV and CM sub-tasks, respectively. Note that the SPF-EER is different from the common EER used in the anti-spoofing community. The difference is that the non-target class is not taken into consideration here but is regarded as the same positive class (bona fide) in the CM community.
The description of EERs can be found in Table~\ref{tab:eers}. The test utterance falls into either of the three classes.
For all of the EERs mentioned above, only the target class is considered positive samples.

\begin{table}[!htbp]
\caption{Three kinds of EERs for evaluation (Adapted from \cite{jung2022sasv}). ``+'' denotes the positive class and ``-'' denotes the negative class. A blank entry denotes classes not used in the metric. SASV-EER is the primary metric for the SASV challenge.
}
\renewcommand{\arraystretch}{1.2}
\centering
\begin{tabular}{c|ccc}
\hline\hline
Evaluation metrics & Target & Non-target & Spoof \\ \hline
SASV-EER           & +      & -          & -     \\ \hline
SV-EER             & +      & -          &       \\
SPF-EER            & +      &            & -    \\
\hline\hline
\end{tabular}
\label{tab:eers}
\end{table}

\begin{table*}[!htbp]
\caption{Comparison of our proposed methods with separate systems and SASV challenge baselines.
}
\renewcommand{\arraystretch}{1.2}
\centering
\begin{tabular}{M{70pt}|R{31pt}R{31pt}R{31pt}R{31pt}|R{31pt}R{31pt}}
\hline\hline
\multirow{2}{*}{\textbf{Systems}} &
  \multicolumn{2}{c}{~\textbf{SV-EER}$\downarrow$} &
  \multicolumn{2}{c|}{~\textbf{SPF-EER}$\downarrow$} &
  \multicolumn{2}{c}{~\textbf{SASV-EER}$\downarrow$} \\ \cline{2-7} 
 &
  \multicolumn{1}{c}{~~~Dev} &
  \multicolumn{1}{c}{~~~Eval} &
  \multicolumn{1}{c}{~~~Dev} &
  \multicolumn{1}{c|}{~~~Eval} &
  \multicolumn{1}{c}{~~~Dev} &
  \multicolumn{1}{c}{~~~Eval} \\ \hline
ECAPA-TDNN     & 1.86  & 1.64  & 20.28 & 30.75 & 17.31 & 23.84 \\
AASIST    & 46.01 &  49.24 &  0.07 &  0.67  &  15.86  & 24.38 \\ \hline
Baseline1 & 32.89 & 35.33 & 0.07  & 0.67  & 13.06 & 19.31 \\
Baseline2 & 7.94 & 9.29 & 0.07  & 0.80  & 3.10  & 5.23  \\ \hline
\texttt{PR-L-I} (Ours) & 2.13 & 2.14 & 0.11  & 0.86  & 1.21  & 1.68  \\
\texttt{PR-S-I} (Ours) & 2.43 & 2.57 & 0.07  & 0.78  & 1.34  & 1.94  \\
\texttt{PR-C-I} (Ours) & 1.95 & 1.64 & 0.97  & 2.94  & \textbf{1.08}  & 2.70  \\
\textbf{\texttt{PR-L-F} (Ours)} & 2.02 & 1.92 & 0.07  & 0.80  & 1.10  & \textbf{1.54}  \\
\textbf{\texttt{PR-S-F} (Ours)}      & 2.02    & 1.94    & 0.07     &     0.80  &  1.10  & \textbf{1.53}   \\
\hline\hline
\end{tabular}
\label{tab:main}
\end{table*}

\subsection{Implementation details}
Our implementation is based on PyTorch~\footnote{Our work is reproducible with code available at \url{https://github.com/yzyouzhang/SASV_PR}.}. The pre-trained embeddings are provided by the SASV organizers. They are extracted with already-trained state-of-the-art ASV and CM systems.
The ASV system is an ECAPA-TDNN~\cite{Desplanques2020} model trained on the VoxCeleb2 dataset~\cite{chung2018voxceleb2}. The CM system is an AASIST~\cite{Jung2022AASIST} model trained on ASVspoof 2019 LA training set~\cite{WANG2020101114}. 
For a speech utterance, the speaker embedding has a dimension of 192 and the CM embedding is a 160-dim vector.

For the Baseline2 model structure, the DNN is composed of four FC layers, each with the number of output dimensions as 256, 128, 64, 2, respectively. Each intermediate layer is followed by a leaky ReLU activation function.
For inference, we use the official trials provided by the SASV challenge organizers as described in Section~\ref{ssec:dataset}. The calibrator in \texttt{PR-C-I} is trained on the bona fide utterances of the development trials.


During training \texttt{PR-L-F} and \texttt{PR-S-F}, we randomly select pairs of utterances from the training set. For the binary cross-entropy loss, we set the prior probability for a target trial as 0.1.
We train our systems using Adam optimizer with an initial learning rate of 0.0003. The batch size is set to 1024. We train the model for 200 epochs and select the best epoch according to the SASV-EER on the development set. The model in the best epoch is used for final evaluation.

\section{Results}
\subsection{Comparison with separate systems and baselines}
To demonstrate the effectiveness of our proposed strategies, we compare our methods with the individual systems and baseline methods in the SASV challenge\footnote{Note that the baseline results we report have differences from those reported in~\cite{jung2022sasv}. Based on our implementation, we achieved close results for ECAPA-TDNN and Baseline1, but better results for Baseline2.
}.
The performance comparison is shown in Table~\ref{tab:main}.

The individual systems perform well on their own tasks but have much worse performance on the other task. The ECAPA model achieves the lowest SV-EER but a high value in SPF-EER. This verifies that the state-of-the-art speaker verification system is vulnerable to spoofing attacks. Quite a number of spoofed trials can deceive the ASV system and degrade the SASV performance. The AASIST system has the lowest SPF-EER but close to 50\% SV-EER. This is reasonable since all bona fide speech, no matter target or non-target, are considered positive samples in training CM systems. The well-trained CM system is not expected to have discrimination ability for ASV.

Both baseline methods surpass the separate systems in terms of SASV-EER, showing the superiority of an ensemble solution for the SASV problem. Baseline1, a score-level fusion-based method, has the same SPF-EER performance as the single CM system but degrades the ASV performance compared to the ECAPA model. This suggests that the non-calibrated scores might degrade the performance on sub-tasks. Baseline2, the embedding level fusion-based model, has much better performance on all three metrics overall with only the SPF-EER degraded a little on the evaluation set. 

All of our proposed systems show a significant improvement over the baseline methods in terms of SASV-EER. 
They also achieve universally good performance over all three metrics. Both the SV-EER and SPF-EER are close to the performance of the best separate model. This shows the effectiveness of our product rule (PR)-based probabilistic framework with our proposed direct inference strategy and fine-tuning method. Our \texttt{PR-S-F} system achieves the best performance on the evaluation trials.

\subsection{Comparison among the proposed strategies}
\label{ssec:comp_pr}
Comparing our proposed systems with direct inference strategy (i.e., with \texttt{-I}) and systems with fine-tuning strategy (i.e., with \texttt{-F}), the latter generally achieve better performance. This suggests the effectiveness of the joint optimization by slacking the conditional independence of ASV and CM subsystems.

Among all the systems with direct inference strategy, we can compare the impact of different choices for the mapping function $f$ applied to the ASV cosine similarity score. The linear mapping achieves better SV-EER and SASV-EER compared to the sigmoid mapping, this might be attributed to the non-linearity of the sigmoid function that distorts the ASV score distribution. The calibrated ASV score achieves the best performance on the development trials in terms of SASV-EER, and the SV-EER is the closest to ECAPA-TDNN, suggesting that the calibration on ASV scores is effective for SASV. However, the calibration degrades the SASV-EER performance and the SPF-EER performance on the evaluation trials prominently. Note that the spoof trials in the development and evaluation trials are generated with different attack algorithms. The performance degradation verifies our hypothesis that the calibration would cause the joint system to overfit the distribution of the trials that the calibrator is trained on hence cannot generalize well to unseen attacks.

Among the two systems with our fine-tuning strategy, both of them achieve top similar performance in all three metrics. This suggests that joint optimization is effective and robust to both linear and sigmoid mapping functions. Although the score mapping functions affect the performance in the direct inference strategy, they do not make much difference in the fine-tuning strategy, thanks to the FC layer re-trained on SASV labels.



\subsection{Ablation study on Baseline1}
\label{ssec:ablation}
Since our model structure is based on Baseline1, we perform an ablation study to recover the components back to the counterparts in Baseline1 and observe the performance degradation. The results are shown in Table~\ref{tab:ablation}.
The performance degradation from \texttt{PR-S-F} to \texttt{PR-S-I} verifies the effectiveness of our proposed joint optimization by fine-tuning.
Both \texttt{PR-S-I} and Baseline1 are direct inference methods. 
Comparing Eq.~\eqref{eq:pr_s} and the formula in Fig.~\ref{fig:baselines} (a), changes on the computation of the SASV score in our proposed approach compared to Baseline1 are: 1) applying sigmoid score mapping on both ASV score and CM score, 2) using multiplication rather than addition. 


\begin{table}[!htbp]
\caption{Results of ablation study from our proposed best performing system \texttt{PR-S-F} to Baseline1.
}
\footnotesize
\renewcommand{\arraystretch}{1.2}
\centering
\begin{tabular}{c|M{22pt} M{22pt}}
\hline\hline
\multirow{2}{*}{Systems} & \multicolumn{2}{c}{SASV-EER}                  \\ \cline{2-3} 
                            & Dev                   & Eval                  \\ \hline
\multicolumn{1}{c|}{\begin{tabular}[c]{@{}c@{}}\texttt{PR-S-F} (Ours) \end{tabular}} & 1.10  & 1.53  \\ \hline
\multicolumn{1}{c|}{\begin{tabular}[c]{@{}c@{}}\texttt{PR-S-I} (Ours) \end{tabular}} & 1.34  & 1.94  \\ 
\multicolumn{1}{l|}{\begin{tabular}[c]{@{}l@{}}Restore multiplication to sum\\ (Baseline1 + score mapping)\end{tabular}}                      & 1.69  & 2.45  \\
\multicolumn{1}{l|}{\begin{tabular}[c]{@{}l@{}}Remove score mapping\\ (Baseline1 + score multiplication)\end{tabular}}                      & 2.16  & 2.89  \\ \hline
\multicolumn{1}{l|}{\begin{tabular}[c]{@{}l@{}}Restore both \\ (Baseline1)\end{tabular}}                                             & 13.06 & 19.31     \\
\hline\hline
\end{tabular}
\label{tab:ablation}
\end{table}

If we change the multiplication back to summation, i.e., $\S_\textit{SASV} = \sigma(\mathcal{S}_\textit{CM}) + \sigma(\mathcal{S}_\textit{ASV})$, the performance degrades to 2.45\% SASV-EER, which is still a relatively good performance. The degradation indicates the superiority of our proposed probabilistic fusion framework with the product rule.

If we only remove the score mapping but keep the multiplication, i.e., $\S_\textit{SASV} = \mathcal{S}_\textit{CM} \times \mathcal{S}_\textit{ASV}$, the performance degrades to 2.89\% SASV-EER, which is also an acceptable performance. 

When we restore both components back to the Baseline1 method, then the SASV-EER performance degrades significantly. This suggests that both components in our proposed \texttt{PR-S-I} make an effective contribution. What exactly causes the dramatic degradation from \texttt{PR-S-I} to Baseline1? Our hypothesis is that the scores output from the ASV and CM subsystems of Baseline1 are in different ranges, and the summation of the scores makes one subsystem dominates the other. Looking at the Table~\ref{tab:main} again, it is the CM system that dominates. Applying score mapping, with multiplication or summation, also addresses this issue. Replacing summation with multiplication, with or without score mapping, addresses this issue, as the difference between the score ranges is just a constant scalar of the final decision score. This explains why both revised methods in Table~\ref{tab:ablation} do not degrade too much from \texttt{PR-S-I}.

In the next section, we will verify this hypothesis by investigating the scores output from the two subsystems of Baseline1, as well as the revised scores after applying score mapping.


\section{Score distribution analysis}

\begin{figure*}[]
  \centering
  \centerline{\includegraphics[width=1\linewidth]{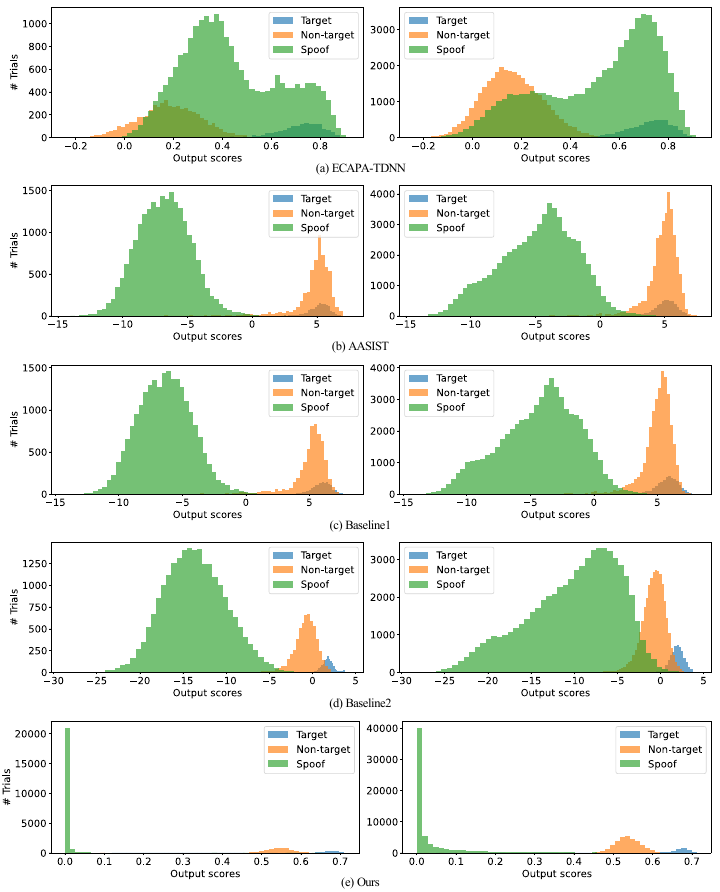}}
\caption{Comparison among score distributions of (a) the ASV subsystem (ECAPA-TDNN), (b) the CM subsystem (AASIST), (c) Baseline1, (d) Baseline2, and (e) our proposed best-performing method \texttt{PR-S-F}. The left column is the performance on the development set and the right column is on the evaluation set. Different colors correspond to the three label classes: target, non-target, and spoof.}

\label{fig:bl_score_dist}
\end{figure*}

Fig.~\ref{fig:bl_score_dist} shows the score distribution of the systems we compared in Table~\ref{tab:main}. We plot the histogram of score distributions on both the official development and evaluation trials.

\begin{figure*}[htbp]
  \centering
  \centerline{\includegraphics[width=1\linewidth]{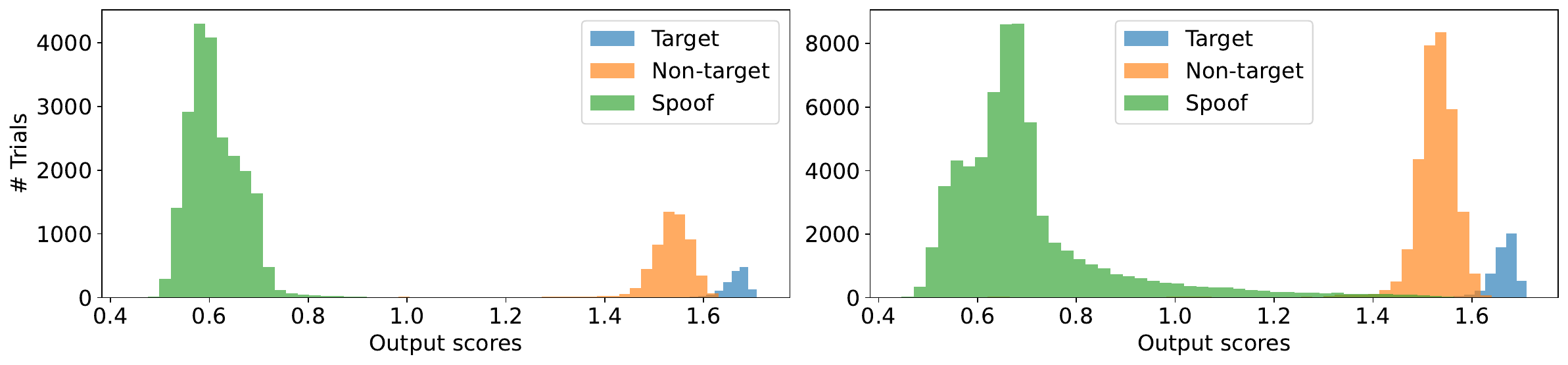}}
\caption{Score distributions of applying score mapping on Baseline1 system.}
\label{fig:abl_score_dist}
\end{figure*}

\begin{figure*}[htbp]
  \centering
  \centerline{\includegraphics[width=1\linewidth]{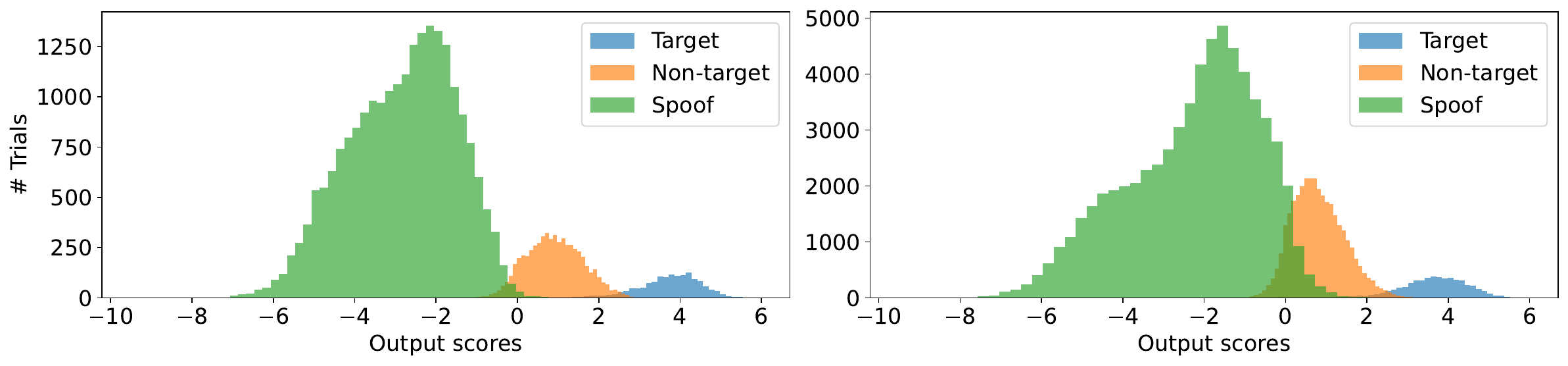}}
\caption{Score distributions of applying score multiplication on Baseline1 system.}
\label{fig:baseline1_mul_score_dist}
\end{figure*}

Fig.~\ref{fig:bl_score_dist} (a) and (b) first plot score distributions of the ASV subsystem (ECAPA-TDNN) and the CM subsystem (AASIST). They demonstrate good discriminative abilities on their individual tasks, but fails to differentiate classes defined in the other task. For example, ECAPA-TDNN well distinguishes \textit{target} and \textit{non-target}, but the distribution of \textit{spoof} expands a wide range, overlapping with both the \textit{target} and \textit{non-target} classes.
This shows that the ASV system is vulnerable to spoofing attacks. It is interesting to see that the scores of spoofing attacks on the evaluation set (right column) are closer to those of the target class. This might suggest that the spoofing attacks in the evaluation set are more challenging to the whole system. 

Similarly, for AASIST in Fig.~\ref{fig:bl_score_dist} (b), the spoof class score is well-separated from the target and non-target classes. However, the target and the non-target classes are highly overlapped since they are both bona fide speech. The CM system only has the ability to discriminate spoofing attacks from bona fide speech. 

For Baseline1 in Fig.~\ref{fig:bl_score_dist} (c), the distribution is similar to that in (b), the difference is that the non-target cluster and the target cluster are deviated by some distance. Recall that Baseline1 takes the sum of the independent scores output by ECAPA-TDNN and AASIST. Comparing (a), (b), and (c), we can infer that the CM system dominates the score. From the score ranges shown in (a) and (b), the absolute values of the CM scores are larger than those of the ASV scores. This verifies our reasoning for why Baseline1 degrades from our proposed \texttt{PR-S-I} so much in the previous section.

For the Baseline2 system in Fig.~\ref{fig:bl_score_dist} (d), the distribution shows that the three classes are more separated than previous systems. 
This suggests that the embedding-level fusion maintains a good discrimination ability for the target class.



From the ablation study in Section~\ref{ssec:ablation}, we find that with simple score mapping and score multiplication, the resulting system is able to achieve a significant improvement over the score-sum baselines. To better understand the mechanisms behind each operation, we plot the histogram of the SASV score distribution with $\S_\textit{SASV} = \sigma(\mathcal{S}_\textit{CM}) + \sigma(\mathcal{S}_\textit{ASV})$ and $\S_\textit{SASV} = \mathcal{S}_\textit{CM} \times \mathcal{S}_\textit{ASV}$ in Fig.~\ref{fig:abl_score_dist} and Fig.~\ref{fig:baseline1_mul_score_dist} respectively. 
From Fig.~\ref{fig:abl_score_dist},  we can observe that the scores are in the range of (0, 2) and the three classes are well separated, indicating the effectiveness of score scaling, where both individual scores are mapped to the same range. 
Similarly, Fig.~\ref{fig:baseline1_mul_score_dist} shows scores from the distinct three classes clearly, but not as well separated as the previous scaling method.


\section{Conclusion}
In this paper, we proposed effective fusion-based methods for spoofing aware speaker verification (SASV). Specifically, we introduced a probabilistic framework with the product rule and a fine-tuning strategy to a score-sum fusion baseline structure. We demonstrated promising performance with a SASV-EER at 1.53\%, a significant improvement from the previous EER of 19.31\%. Our ablation study verified the effectiveness of our proposed strategies and we investigated the SASV decision score distributions of various systems.

\section{Acknowledgment}
This work is supported by National Science Foundation grant No. 1741472, New York State Center of Excellence in Data Science award, and funding from Voice Biometrics Group. You Zhang thanks the synergistic activities provided by the NRT program on AR/VR funded by NSF grant DGE-1922591. 

The authors would like to thank Xinhui Chen for delivering a literature review presentation on \textit{Joint Speaker Verification and Spoofing Countermeasure Systems} during her master's study at University of Rochester.

The authors would like to thank the organizers of the SASV 2022 challenge for providing the pre-trained embeddings.

\bibliographystyle{IEEEbib}
\bibliography{Odyssey2022_BibEntries}

\end{document}